\documentstyle[prd,aps,epsf]{revtex} 
\begin{document}            
\draft

\def\gsim{ \lower .75ex \hbox{$\sim$} \llap{\raise .27ex \hbox{$>$}} }
\def\lsim{ \lower .75ex \hbox{$\sim$} \llap{\raise .27ex \hbox{$<$}} }

\twocolumn[\hsize\textwidth\columnwidth\hsize\csname 
@twocolumnfalse\endcsname

\title{Dark matter in Draco and the Local Group: \\ 
Implications for direct detection experiments}

\author{Ben Moore\footnotemark[1] 
and Carlos Calc\'{a}neo--Rold\'{a}n}
\address{Department of Physics, Durham University, 
Science laboratories, Durham, DH1 3LE, UK.}

\author{Joachim Stadel, Tom Quinn, George Lake and Sebastiano Ghigna}
\address{Department of Astronomy, University of Washington, 
Seattle, WA 98195, USA}

\author{Fabio Governato}
\address{Osservatorio Astronomico di Brera, via Bianchi 46, Merate, Italy}

\date{\today}
\maketitle

\begin{abstract}
We use a cosmological simulation of the Local Group to make
quantitative and speculative predictions for direct detection 
experiments.
Cold dark matter (CDM) halos form via a complex series of
mergers, accretion events and violent 
relaxation which precludes the formation of significant caustic 
features predicted by axially symmetric collapse.  
The halo density profiles are combined with 
observational constraints on the galactic mass distribution to 
constrain the local density of cold dark matter to lie in the range 
$0.18 \ \lsim \ \rho_{_{CDM}}(R_\odot)/{(\rm GeV\ cm^{-3})} \ \lsim \ 0.30$. 
In velocity space, 
coherent streams of dark matter from tidally disrupted halos fill the 
halo and provide a tracer of the merging hierarchy. 
The particle velocities within 
triaxial CDM halos cannot be approximated by a simple Maxwellian distribution
and is radially biased at the solar position.
The detailed phase space 
structure within the solar system will depend on the early merger 
history of the progenitor halos and the importance of major mergers 
over accretion dominated growth.   
We follow the formation of a 
``Draco'' sized dSph halo of mass $10^8M_\odot$ with several million 
particles and high force accuracy.  
Its internal structure and 
substructure resembles that of galactic or cluster mass halos: the 
density profile has a singular central cusp and it contains 
thousands of sub-halos orbiting within its virial radius demonstrating 
a self-similar nature to collisionless dark matter sub-clustering. 
The singular cores of substructure halos always survive 
complete tidal disruption although mass loss is continuous and rapid.
Extrapolating wildly to earth mass halos 
with velocity dispersion of 1 m s$^{-1}$ (roughly equal to the free streaming 
scale for neutralinos) we find that most of the dark matter may remain 
attached to bound subhalos.  Further numerical and analytic work is 
required to confirm the existence of a detectable smooth component.
\end{abstract}

]

\footnotetext[1]{Email:ben.moore@durham.ac.uk}

\section{Introduction}
\label{s:intro}

Revealing the nature of dark matter is fundamental to cosmology  and
particle physics. A combination of observations and theory suggests
that the dark matter consists of non-baryonic particles and in the
large class of hierarchical cosmogonic models, a universe with a
matter density  dominated by cold dark matter (CDM) remains plausible
albeit with some potential problems on small scales  (c.f., Refs.
~\cite{moore94,flores94,burkert95,klypin99a,moore99a,sellwood00}).

The only convincing method for confirming the existence and nature 
of dark matter is by direct detection.  There are two prime candidate
particles. Neutralinos are the lightest super-symmetric particles,
otherwise known as WIMPS (weakly interacting massive particles). These
can be identified in a laboratory by looking for phonons or a
temperature increase from elastic scattering and nuclei recoil in
various materials.  Neutralinos also have a small cross-section for
annihilation into decay products such as high energy photons or
neutrinos that may be identified indirectly as a diffuse
background from the galactic halo ~\cite{berg98}. 
Axions are another type of
hypothetical particle that have the additional motivation of ensuring
that the strong interaction conserves charge-parity; these can be
identified by stimulating their conversion to photons within a
magnetic cavity (c.f., Ref. ~\cite{griest88}).

Many laboratory and some space based experiments utilising these
detection methods are in progress and after a great deal of
technological development they are beginning to probe the parameter
space allowed by cosmological and particle physics constraints
(e.g. Refs. ~\cite{DRIFT99,DAMA00,CDMS00,EDELWEISS}). These experiments are
all highly sensitive to the local density of particles and their
velocity distribution ~\cite{copi01,ullio00,widrow00,stiff01}.  
The flux of gamma-rays on Earth from neutralino
annihilations in the galactic halo is also very sensitive to any
substructure in the dark matter ~\cite{calca00}. 
It is therefore crucial to understand the phase space structure
of galactic halos in the hierarchical CDM model in order that
experiments can be fine tuned to search for the appropriate signals
and that potential signals can be interpreted.

Many of the ongoing direct detection experiments adopt the principle
that CDM particles passing near Earth have a smooth continuous density
distribution with an isotropic Maxwellian velocity distribution with
3-d dispersion $\sim 270{\rm ~km~s^{-1}}$.  We shall demonstrate that these
assumptions are incorrect.  Recent theoretical work has examined the
possibility of velocity anisotropies resulting from halo rotation
~\cite{brhlik99} or triaxiality  ~\cite{kamio98,evans00}. Other
halo models have also been studied e.g., Ref. ~\cite {sikivie99}, 
who assumes
axially symmetric and cold collapse of matter to infer the presence of
caustic rings in the solar neighbourhood.  However, two decades of
cosmological simulations of the cold dark matter model has clearly
demonstrated that dark matter halos form via a series of mergers and
accretions of dark matter clumps along highly filamentary mass
distributions.  Assuming symmetry and
locally cold flows is an incorrect over-simplification of the true
hierarchical growth ~\cite{moore01}.

At a given point in a CDM halo, the ``smooth'' dark matter background
arises from material that has been tidally stripped from less massive
halos (e.g., Ref.~\cite{ghigna98}). The velocity distribution of particles
reflects the mass distribution of progenitor halos that have merged
and accreted into the final system.  The power spectrum of
fluctuations in the CDM model allows small dense dark matter halos to
collapse at very early epochs. The cutoff scale from the free
streaming of neutralinos is approximately $10^{-12}M_\odot$ 
~\cite{schwarz00}, although it is possible that the QCD transition may
introduce features in the power spectrum that allow clumping of CDM on
even smaller scales ~\cite{schmid97}.

Previous simulations of the non-linear growth of halos systematically
found triaxial structures with a smooth internal structure ~\cite{white76}. 
It was thought that the merging and
virialisation process would naturally destroy the precursor halos,
erasing the substructure and any memory of the initial conditions 
~\cite{white78}.  However, the over-merging of dark matter
substructure halos was due primarily to poor force and mass resolution
~\cite{moore96}.  After several decades of code development, matched
with increased hardware performance, numerical simulations have
finally achieved a resolution of sub-kpc scales within a cosmological
context. Recent high resolution numerical simulations can follow the
evolution and survival of dark matter substructure halos (subhalos) as
they orbit within dense environments ~\cite{klypin99a,ghigna98,moore98}.

We have previously used numerical simulations to investigate the 
structure of dark matter halos across a range of mass scales from 
$10^{12}M_\odot$ to $10^{15}M_\odot$. Rather than run many 
simulations at low resolution we have simulated several at the 
highest resolution possible with current technology.  One of the 
key results that we have found
is that substructure within hierarchical models is nearly scale
free. The distribution and orbital properties of
``halos-within-halos'' is independent of halo mass \-- CDM galaxy
halos contain literally thousands of halos more massive than those
that surround the dwarf satellite galaxies in our own halo, Draco and
Ursa-Minor ~\cite{moore99b,klypin99b}.  In this paper we
explore the phase space distribution of CDM within galactic and
sub-galactic halos addressing questions of direct relevance to the
direct detection of dark matter. Given our limited resolution
compared to laboratory scales, we discuss the
extrapolation of our results to earth mass and solar system scales.

Section II discusses the simulation techniques and results of a high
resolution simulation of the Local Group and a dark matter mini-halo.
Section III discusses the observational constraints on the dark matter
halo surrounding the Milky Way in the context of the CDM hierarchical
model.  Sections IV and V examine the survival of dark matter
substructure and the velocity distribution of dark matter at the sun's
position within the Galactic halo.  Section VI summarises the main
results of this paper.

\section{A cosmological simulation of the Local Group} 

Initially we perform a simulation of a 50 Mpc cube of a standard CDM
universe with $\Omega_{_{CDM}}=1$, normalised such that $\sigma_8=0.7$
and the shape parameter $\Gamma=0.5$ (a Hubble constant of 100
km s$^{-1}$ Mpc$^{-1}$ is adopted throughout).  Although we use
simulations within the context of the standard cold dark matter model,
our results do not depend sensitively on the mass density of CDM or on
the presence of a cosmological constant. 
The parent simulation contains
$144^3$ particles with a mass of $1\times10^{10}M_\odot$ and we use a
force softening of 30 kpc. This is adequate for resolving the location
and global properties of Galactic mass dark matter halos. The particle
distribution is evolved using the extensively tested parallel
treecode, ``PKDGRAV'', that has accurate force resolution, periodic
boundaries and variable timesteps.  The code uses a co-moving spline
softening length such that the force is completely Newtonian at twice
our quoted softening lengths.

At a redshift z=0 we search for a pair of dark matter halos with
similar dynamical properties as the Local Group, including mass,
separation, infall velocity  and proximity to a large Virgo-like
cluster. The cosmological volume contains many binary halos of similar
mass as the Local Group, but only a few that are in a similar dynamical 
state ~\cite{governato97}. The particles that define one
of the low resolution Local Group candidates (those particles within
twice the virial radii of both dark matter halos) are traced back to a
redshift z=100 to identify a volume that collapses into the final
system.  Within this arbitrary shaped region we extrapolate the power
spectrum to smaller scales, matched at the boundaries such that both
the power and waves of the new density field are identical in the
region of overlap. This region is populated with a new subset of less
massive particles and zones of heavier particles are placed outside
this region, which allows the correct tidal field to be modelled from
the entire cosmological volume of the initial box.

\twocolumn[\hsize\textwidth\columnwidth\hsize\csname 
@twocolumnfalse\endcsname
\begin{figure}
\begin{center}
\centering
\epsfysize=7.4truein
\epsffile{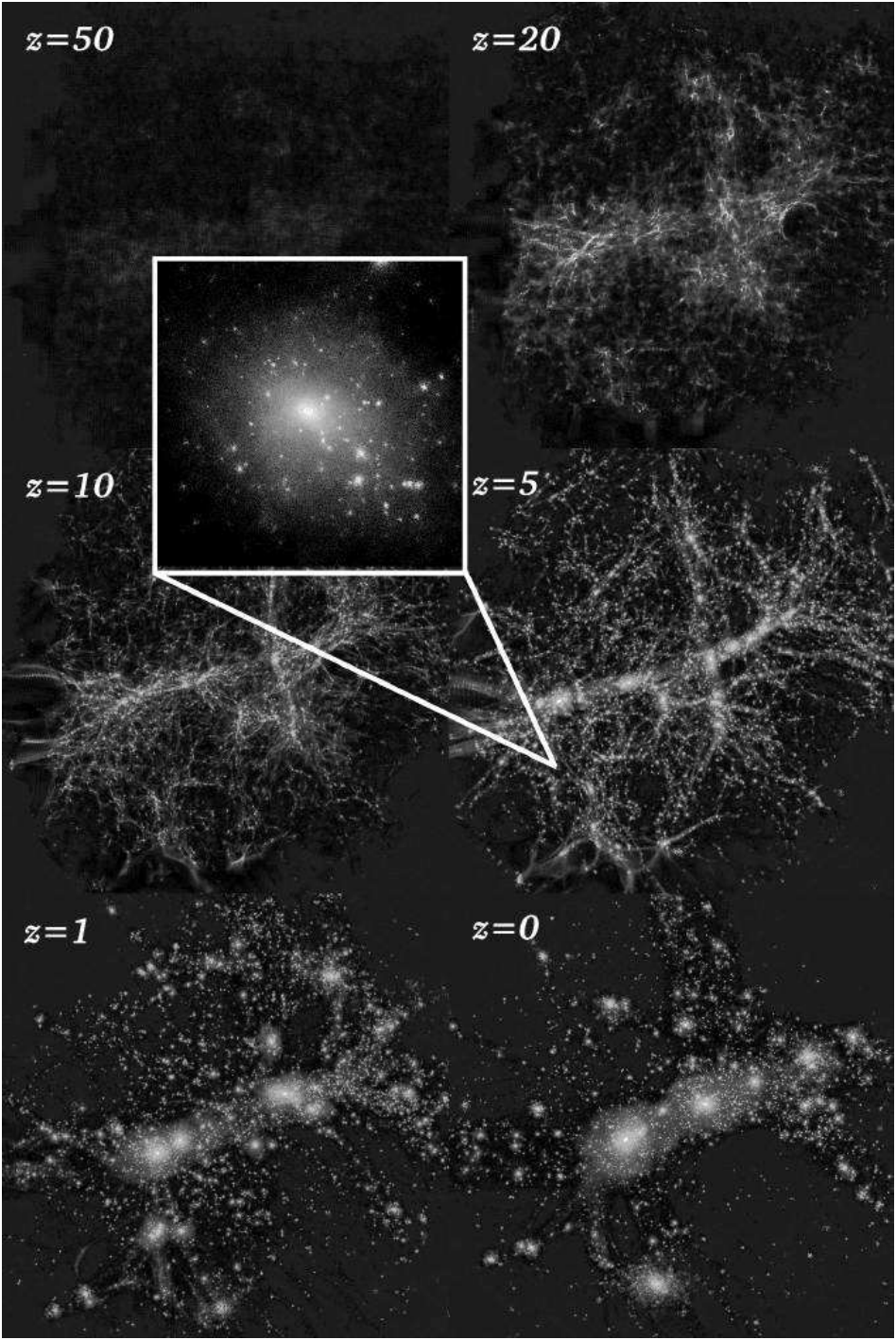}
\end{center}
\caption{The hierarchical formation of a ``Local Group'' within a cold 
dark matter universe.  The grey scale shows the local density of dark 
matter at the indicated redshifts. At a redshift $z=0$ we have two dark 
matter halos separated by 1 Mpc infalling at $\approx 100$ km s$^{-1}$ 
with circular velocities  $\approx 200$ km s$^{-1}$.  The inset box 
at z=5 shows the internal structure of a dark matter subhalo with total 
mass $10^8M_\odot$. The region that collapses into this object was 
followed at a higher resolution such more than that $10^6$ particles 
lie within its virial radius of 2 kpc. (The size of each panel is 4 
comoving Mpc and the inset panel is 4 kpc.)}
\label{f:001}
\end{figure} ]

\begin{figure}[h]
\centering
\epsfxsize=\hsize\epsffile{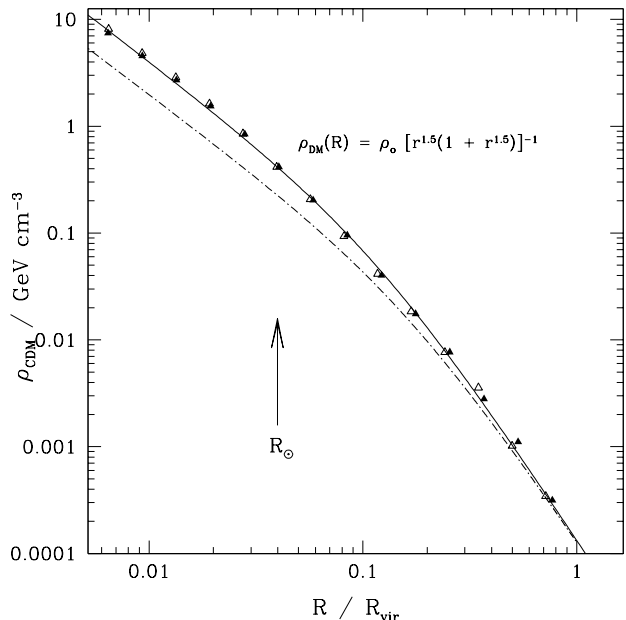}
\caption{The density profiles of the two high
resolution ``Local Group'' CDM galactic halos.  The density profiles of
virialised equilibrium CDM halos are well fitted by the indicated
curve, where $r=R/(R_{vir}/c_{moore})$.  The density at the solar
radius $R_\odot$ depends on the cosmological model adopted, the
standard CDM model simulated here would give values of
$\rho_{_{CDM}}(R_\odot)$ that are approximately a factor of two higher
than a $\Lambda$ dominated CDM model (dot--dashed curve).}
\label{f:002}
\end{figure}

The particle mass in the high resolution regions is $1 \times
10^6M_\odot$ and the spline force softening is set to 0.5 kpc.  The
starting redshift is increased such that the initial fluctuations are
less than one percent of the mean density and we then re-run the
simulation to the present epoch. This calculation required $\sim
150,000$ T3E cpu hours to evolve to the present day. The particles on
the shortest timestep require more than 100,000 individual steps and in
total we calculated $\sim 10^{15}$ floating point operations.  The
final virial radii and locations of the halos agree with those in 
the lower resolution simulation to within a few percent.
(For reference, the accuracy parameters we use with PKDGRAV are 
$\theta(z>2)=0.5$,
$\theta(z<2)=0.7$,
$\epsilon=0.2$.)
At the final
time, each halo contains $\sim 2\times 10^6$ particles and is well
resolved at a distance of 1\% of the virial radius, $r_{vir}\approx
200$ kpc.  Snapshots of the formation of this pair of halos is
shown in Fig.~\ref{f:001}and their spherically averaged density profiles are
plotted in Fig.~\ref{f:002}.

A good fit to the density profiles of CDM Galactic halos is given by
$\rho(r) \propto 1/[(r/r_s)^{1.5}+(r/r_s)^3]$ where the scale  length
$r_s=r_{vir}/c_{moore}$ and the concentration  $c_{moore} \approx 0.67
c_{nfw}$. Figure~\ref{f:002} shows the density profile expected in the
concordance $\Lambda$CDM model in which the concentration is about a
factor of two lower than the standard CDM model simulate here.

\subsection{The fractal nature of dark matter}

The smallest scale on which we have evidence for dark matter is within
the dwarf spheroidal satellites that orbit within the Galactic
halo. Our Local Group simulation contains over 2000 of these systems
with peak circular velocities, $v_{peak}$, larger than 10 km s$^{-1}$ and
most of these orbit  within the virial radius of the two large
halos. At our current resolution, most of the subhalos are smooth i.e.
``halos within halos within halos'' are found only within the largest
substructures.  If we could increase the resolution to $10^{10}$
particles we could accurately study the internal structure of dark
matter subhalos and investigate the clustering of matter on very small
scales. Such a simulation is probably about a decade away, therefore we shall
repeat the  ``volume renormalization'' process described above on a
single dark matter subhalo from the Local Group simulation.

First we identify a dark matter mini-halo with a velocity dispersion
of 10 km s$^{-1}$ that accretes into one of the Local Group 
binary halos at a late epoch, $z\approx 3$.  This halo would resemble
the dark matter dominated dwarf spheroidal galaxies that orbit within
the Galactic halo. 
The particles from the low resolution halo
are traced back to their starting positions to identify the
region to be resimulated at higher resolution. We start this re-simulation
at a redshift z=150 and our particle mass is $50M_\odot$ and our force
softening is 10 parsecs -- we can resolve dark matter halos collapsing
at redshifts $z=50$ with characteristic velocities $v_{circ}\lsim 1$
km s$^{-1}$. The ``Draco'' mass halo collapses/virialises at a redshift
$z=7$ and by $z=3$ we have to stop the simulation since it
becomes contaminated with heavy particles from the bounding regions as
it enters the less well resolved galactic mass halo.

\begin{figure}[h]
\centering
\epsfxsize=\hsize\epsffile{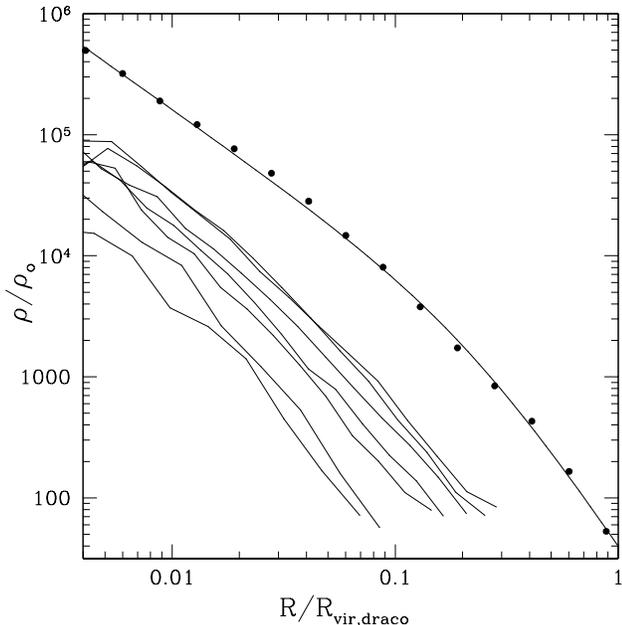}
\caption{The density profile of the ``Draco'' dark matter mini-halo at z=4
  (dots) and seven randomly selected progenitor halos at z=15.  
The solid curve shows the best fit density profile to the Draco halo, 
constrained to have an outer slope of -3 but leaving the
inner slope as a free parameter, resulting in a value of -1.3.}
\label{f:003}
\end{figure}

The blown up region in Fig.~\ref{f:001} shows the mass distribution within the
virialised extent of the Draco halo at $z=5$. At this epoch the
mini-halo has a mass of $10^8M_\odot$, circular velocity of 15 km s$^{-1}$
and virial radius of 2 kpc  in physical coordinates. Roughly 10\% of
its mass is attached to over 600 dark matter substructure halos
containing more than 32 bound particles.  The characteristic slope of
the power spectrum on this scale is $n_{eff}\approx -2.9$ so that all mass
scales are turning non-linear and collapsing at a similar
time. In order for subhalos to survive,
the timescale between collapse and accretion into a
larger object must be sufficiently long to allow the halos to virialise and
form dense cores. The density profiles of progenitor halos at z=15 shown
in Fig.~\ref{f:003} demonstrate that they also have singular cuspy 
inner profiles.

\begin{figure}
\centering
\epsfxsize=\hsize\epsffile{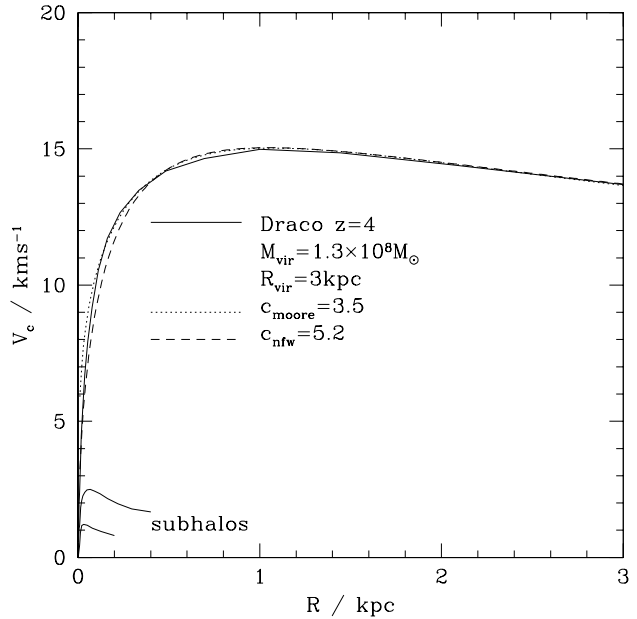}
\caption{The circular velocity curve of the ``Draco'' dark matter mini-halo  
at z=4 (solid curve). The dotted and dashed curves are the best fit 
Moore {\it et al.} and NFW density profiles respectively. The circular 
velocity curves of two small subhalos with circular velocities $\approx 1$ 
km s$^{-1}$ are also plotted.}
\label{f:004}
\end{figure}

Figure~\ref{f:003} shows the density profile of the Draco satellite at z=4. 
In physical
coordinates the rotation curve peaks at 15 km s$^{-1}$ at $r_{peak}=1$ kpc and
$r_{vir}=3$ kpc. We also plot the best fit Moore {\it et al.} 
~\cite{moore99b} and NFW density profiles that have concentrations 
$c_{moore}=3.5$ and $c_{nfw}=5.2$. This latter value is close to that 
predicted by Navarro {\it et al.} ~\cite{navarro96} (NFW) scaling for an
object of this mass $1.3\times 10^8M_\odot$ at this redshift. In fact, the 
NFW scaling provides a good fit to the smallest well resolved progenitor 
halos with circular velocities $\sim 1$ km s$^{-1}$ that also have cuspy 
singular density profiles (c.f., Ref. ~\cite{jang01}). Figure~\ref{f:004} 
shows the circular velocity profile of the Draco halo as well as 
for two of its tiny subhalos and the fits shown in Fig.~\ref{f:003}.

\begin{figure}
\centering
\epsfxsize=\hsize\epsffile{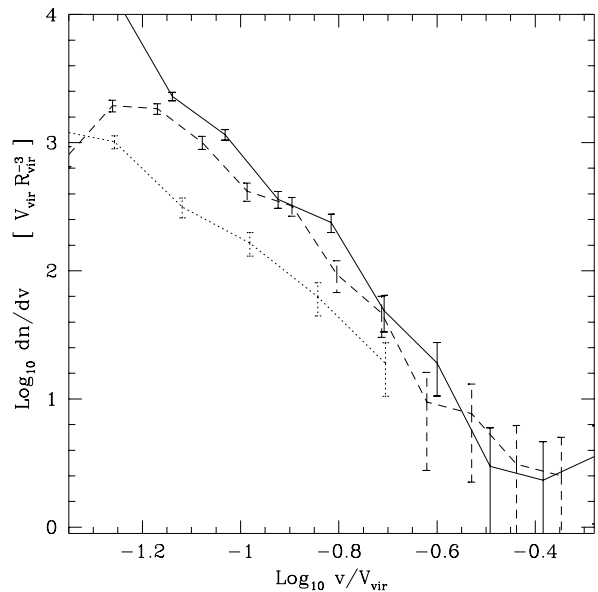}
\caption{The velocity distribution functions
for dark matter subhalos within a CDM cluster, galaxy and mini-halo. Each
halo has been simulated with $\gsim 10^6$ particles and using
softening lengths $\lsim 0.002R_{vir}$. The peak velocities of the
subhalos are normalised by the circular velocity at the virial radius
of their parent halos.}
\label{f:005}
\end{figure}

We identify subhalos within the Draco halo using the same procedures
as in Ref. ~\cite{ghigna98}.  The circular velocity function of subhalos
within Draco, Galactic and Cluster mass CDM halos from the same
cosmological model are compared in Fig.~\ref{f:005}.  The masses of these
objects vary by 7 orders of magnitude yet they all contain similar
amounts of dark matter substructure.  The slope of the circular
velocity function within the ``Draco'' halo, ``Galaxy'' halo and
``Virgo'' halos are $n(v_c)\propto v_c^{-\alpha}$ with $\alpha=3.0,
3.7$ and 3.9 respectively.

The dark matter within virialised systems is nearly self-similar or
fractal-like.  The slope of the circular velocity function within the
mini-halo is shallower than the galaxy and cluster halos which may
reflect the fact that we are approaching the assymptotic ``n=-3'' part
of the power spectrum. (The amplitude of the curve is also lower, due in
part to the redshift at which we are studying the halo -- a halo of
the same circular velocity today would have a virial radius 5 times larger.)
We also caution that this halo has a very particular merger history that may
not be typical of other halos of this mass.

\section{The mean density of dark matter at $R_\odot$}

The flux of particles through a detector depends on the local density
of dark matter and the velocity distribution of particles relative to
the earth's motion through the galaxy ~\cite{freese88}. 
Here we combine several
observational constraints on the baryonic distribution within 
the Milky Way with our numerical simulations of halo structure 
to determine the possible range of CDM halos that may surround the Galaxy.

The density profile of CDM halos within a specific cosmology follow a
single parameter family uniquely determined by their mass. 
The cosmological scatter of halo concentrations is about 30\% for virialised
halos not undergoing current mergers ~\cite{navarro96,eke01}.
This scatter is largely due to 
variations in the virial radius at a fixed circular velocity and is sensitive
to the presence of large substructure halos. If
we consider just the structure internal to $v_{peak}$ then deviations from
our density profile are only of the order 10\% (c.f. Figure 1 of 
Ref. ~\cite{moore99b}). The scatter in central density profiles is small 
because this region is completely relaxed and rarely contains substructure 
halos.

\subsection{Upper limit}

A single observed value of the circular velocity is insufficient to constrain
the properties of the Galactic CDM halo, although the most
important quantity to measure is the circular velocity at the sun's
position. Unfortunately, the structural properties of the Galaxy are
not well known, {\it e.g.} the solar position $7<R_\odot/{\rm
kpc}<8.5$, the disk scale length $2<r_d/{\rm kpc}<4$, the local
circular velocity $190<v_{\odot}/{\rm km s^{-1}}<230$ ~\cite{sackett97}.  
The maximum
density of CDM at $R_\odot$ can be determined by calculating the minimum
contribution to $v_\odot$ from the combined baryonic components.

The mass of an exponential disk is $M_d=2\pi
r_d^2\Sigma(R_\odot) e^{R_\odot/r_d}$ where $\Sigma(R_\odot)$ is the
vertical column density of baryons at the solar position
$R_\odot$.  Direct observations of the local stellar and gaseous distributions
or dynamical estimates of the gravitating mass in the disk yield values
of $\Sigma(R_\odot)$ lie in the range 40--90$M_\odot/{\rm pc}^2$
~\cite{kuijken91,flynn94,flynn99,dehnen98}.
Thus the baryonic disk mass lies in the range $4-8\times 10^{10}M_\odot$, 
of which $67\%$ lies within $R_\odot$. The Galactic bulge 
contributes a further mass of $1-2\times 10^{10}M_\odot$ ~\cite{gerhard99}.

\begin{figure}
\centering
\epsfxsize=\hsize\epsffile{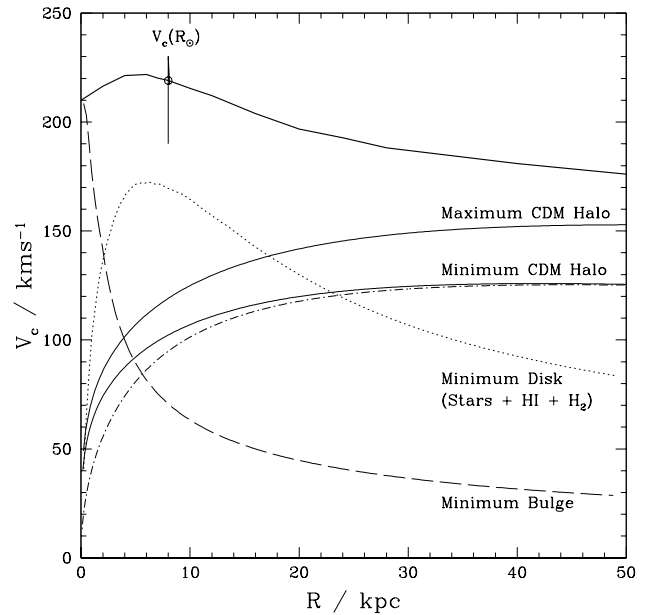}
\caption{The dashed and dotted curves show the minimum
contribution to the Galactic circular velocity curve from bulge and
disk components respectively.  The solid curves show the minimum and
maximum allowed CDM halos with central density cusp 
$\rho(r) \propto r^{-1.5}$.  
The upper thick solid curve is the total circular velocity
profile of the Galaxy for the case in which the CDM halo is the
maximum allowed by observational constraints. The minimum CDM halo is
the least massive halo that can cool the observed mass of baryons
within a Hubble time. The dot--dashed curve is a minimum ``NFW'' halo
with central cusp $\rho(r) \propto r^{-1}$.} 
\label{f:006}
\end{figure}

How does this compare to independent measurements? In order to explain the
microlensing optical depth of K-giant stars towards the Galactic bulge, the
stellar mass within $R_\odot$ must be $>7.6\times 10^{10}M_\odot$ 
~\cite{binney00}.  This is close to the mass of stars that a maximum
disk allows and may be evidence that the Milky Way is a barred galaxy
~\cite{gerhard00}. 
A further constraint on the central dark matter density can be derived
using the kinematics of barred galaxies. The existence of rapidly
rotating bars and the strength and position of shocks in their gas
flows, both indicate low central dark matter densities 
~\cite{debatt98,weiner01}.   The analyses of these authors constrain
the ratio $(v_{disk}/v_{halo})^2>2$ measured at $2r_d$ at the
$2\sigma$ level.  Applying these constraints to the Milky Way with
$v_\odot=220$ km s$^{-1}$ and $v_{bulge}(R_\odot)=70$ km s$^{-1}$, implies 
circular
velocities at $2r_d$ of $v_{halo}<120$ km s$^{-1}$ and $v_{disk}>170$
km s$^{-1}$. These parameters for the disk lie in the range discussed 
above and are illustrated in Fig.~\ref{f:006} as our fiducial 
Galactic model.

Summarising these constraints allows us to estimate the maximum mass
of dark matter within $R_\odot$, given that we do not want to
overestimate the contribution to the observed circular velocity.
This gives $M_{_{CDM}}\lsim 3\times 10^{10}M_\odot$ for 
$v_\odot=220$ km s$^{-1}$.
Adopting the currently favoured $\Lambda$CDM model with
$\Omega_\Lambda=0.7$, $\Omega_{CDM}=0.3$, $\sigma_8=0.9$, $h=0.65$ 
constrains the structure of the maximum CDM
halo to be; $c_{moore}=6$ ($c_{nfw}=9$), $v_{peak}=153$ km s$^{-1}$,
$R_{vir}=250$ kpc, $v_{vir}=124$ km s$^{-1}$,
$M_{vir}=9\times10^{11}M_\odot$, $R(v_{peak})=52$ kpc. 
Using the halo density profile defined in Section II
we find a mean density of dark matter 
$\rho_{_{CDM}}(R_\odot) \ \lsim \ 0.23 \ {(\rm GeV\ cm^{-3})}$.
(Adopting an NFW profile with inner cusp of -1 would lower this value by
about 15\%.) For $v_\odot=230$ km s$^{-1}$ this density could rise
by 30\% to
$\rho_{_{CDM}}(R_\odot) \ = \ 0.3 \ {(\rm GeV\ cm^{-3})}$.

\subsection{Lower limit}

A lower limit to the mass of the Galactic halo can be found assuming
that it must be massive enough to ``cool'' the observed quantity of
baryons to form the visible galaxy ~\cite{white93,navarro00}. 
This assumes that those baryons initially outside
$r_{vir}$ have not had time to cool and reach the central disk. This
is probably an underestimate of the halo mass since a large fraction
of the baryons may be external to the disk in a warm component --
perhaps ejected via super novae feedback which is an essential
ingredient to galaxy formation in the CDM model.

To estimate a lower limit to the halo mass we take a baryon fraction
given by the upper limit from nucleosynthesis, $\Omega_b=0.019h^{-2}$ 
~\cite{tytler99} 
and a low Hubble constant $h=0.6$.  For a total baryonic mass
of $8\times 10^{10}M_\odot$, the halo mass must be
$M_{vir}>(\Omega_o/\Omega_b)M_{baryon} > 4.6\times 10^{11}M_\odot$ in
order to cool the observed amount of baryons.  (This calculation
limits the total amount of ejected baryons to be less than the mass
that currently resides in the Galaxy.)  Within our adopted
$\Lambda$CDM model this minimum CDM halo would have a density profile
with parameters; $c_{moore}=6$, $R_{vir}=200$ kpc, $v_{vir}=100$ km s$^{-1}$,
$v_{peak}=125$ km s$^{-1}$, $R(v_{peak})=40$ kpc and
$M_{vir}=4.6\times10^{11}M_\odot$.
This leads to a lower limit of
$\rho_{_{CDM}}(R_\odot) \ \gsim \ 0.18 \ {(\rm GeV\ cm^{-3})}$.

Although the Galactic CDM halo is tightly constrained we find that 
a $\Lambda$CDM halo is compatible with the observational constraints.
Our fiducial Galactic model is also consistent with the total mass
inferred from the orbits of its satellites ~\cite{wilkinson99} 
and the value of the circular
velocity $v_c(50{\rm kpc})\approx 200$ km s$^{-1}$ found
from modelling the Sagittarius tidal stream ~\cite{ibata00}.  
Detailed modelling of the thickness of the Galactic
gas disk provides another constraint on the local dark matter 
density~\cite{olling2001}. Finally, we note that
it remains to be resolved whether or not the adiabatic
contraction from the cooling baryons would increase the central
CDM density beyond that allowed by the observations ~\cite{barnes84}.

\section{Structure in density space}

The previous calculation of the dark matter density assumes that the
particles are smoothly distributed.  If the dark matter is physically
clustered on small scales then this estimate could radically change.
For example, if the first objects that collapse in the CDM hierarchy
are small and dense enough, they will survive the Galaxy's tidal field
and remain bound.  The chances of the earth moving through one of
these clumps may be so small that direct detection would not succeed.
Here we discuss the tidally limited structure of CDM halos within halos.

The first numerical studies of collisionless dark matter halo
collapses were carried out by Peebles ~\cite{peebles70} and White 
~\cite{white76}. 
Although the final virialised structure had global mass and size
comparable to observed structures, all traces of substructure
had vanished. Named the ``overmerging'' problem, small dark matter
structures were initially resolved in the simulations but disrupted as
soon as they merged into a more massive structure. This turned out to
be a resolution problem ~\cite{moore96}, and as
demonstrated in Section II, we can now resolve many thousands of dark
matter substructure halos orbiting within the virialised extent of
Galactic and sub-Galactic halos.

\begin{figure}
\centering
\epsfxsize=\hsize\epsffile{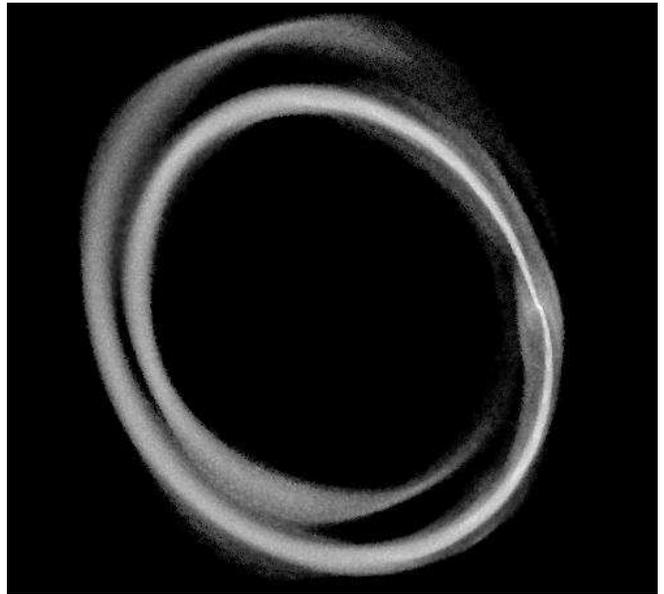}
\caption{The evolution of a dark matter subhalo orbiting on a circular 
polar orbit within  a Galactic potential at 20 kpc.  The snapshot is 
after 3 Gyrs, about 5 orbits. The tidal streams of dark matter 
lead and trail the surviving dark matter clump of which just 0.3 
percent of the initial mass remains bound.}
\label{f:007}
\end{figure}

However, even our highest resolution simulations show little
substructure within about $10$\% of the virial radius. This is
primarily a resolution effect since we do not have enough particles to
resolve the cores of tidally stripped halos. 
CDM halos have singular cuspy profiles on all mass scales
simulated to-date from $10^6M_\odot$ to $10^{15}M_\odot$, therefore a
central core will always survive but would require extremely
high mass resolution  to resolve it.

If the number density of subhalos continues as a power law to 
very small masses would we expect a smooth component of matter at the
solar radius?  This question is quite subtle since the surviving
fraction of halos at a given radius depends also on how the parent
halo is assembled. i.e.  is the formation dominated by major (similar
mass) mergers or by accretion driven growth? In the former case halos
will always be smooth since the halo centers coalesce to form a single
system. In the latter case, halo cores always survive but may not have
much mass attached to them. Further simulations and  analytic modeling
are required to fully address this question, here we  make an estimate
of the radius at which accreted  subhalos can retain most of their
mass intact from the Galactic tidal field.

Although it is physically impossible to completely disrupt a CDM
subhalo, once the tidal radius, $r_t$, imposed by the Galaxy
approaches the radius at which the density profile becomes shallower
than -2, the subhalo loses mass  rapidly.  If $r_t> r_{peak}$ then the
halo will be stable and lose very little mass. This behaviour can be
understood in terms of the energy distribution of particles compared
with the escape velocity at different radii.  This process is
illustrated in Fig.~\ref{f:007} in which we examine tidal mass loss for the
extreme case.  We construct an equilibrium dark matter halo with
a peak circular velocity of 10 km s$^{-1}$ using $10^7$ particles, force
softening of 10 parsecs and isotropic particle orbits. The model is an
equilibrium Hernquist profile that has an inner density profile
$\rho(r)\propto r^{-1}$ and is  essentially a equilibrium replica of
the Draco halo simulated earlier.

\begin{figure}
\centering
\epsfxsize=\hsize\epsffile{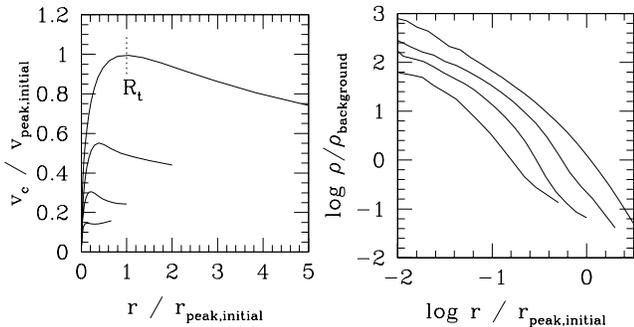}
\caption{The evolution of the circular velocity and density profiles of 
the satellite shown in Fig.~\ref{f:007}. The curves show the initial 
configuration and subsequent times of 1, 2 and 3 Gyrs. The satellite is 
on a circular orbit at 20 kpc from the center of a Galactic potential and 
the theoretical tidal radius for this orbit is indicated.}
\label{f:008}
\end{figure}

\begin{figure}
\centering
\epsfxsize=\hsize\epsffile{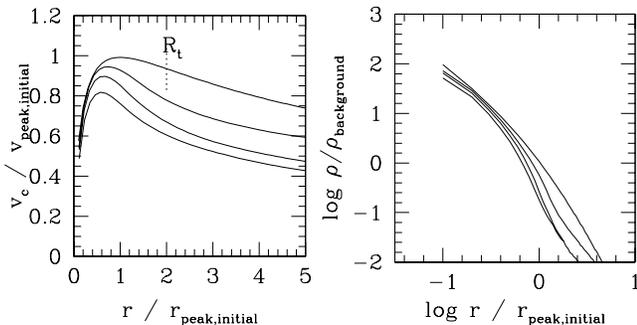}
\caption{The evolution of the circular velocity and density profiles of 
the satellite
shown in Fig.~\ref{f:007}. The curves show the initial configuration and
subsequent times of 1, 2 and 3 Gyrs. The satellite is on a circular
orbit at 40 kpc from the center of a Galactic potential.}
\label{f:009}
\end{figure}

Figure~\ref{f:007} shows the evolution of this model placed on a circular 
orbit
within a Galactic potential for three billion years.  The model has
been constructed such that the tidal radius imposed by the Milky Way
is equal to the radius at which the satellite's circular velocity
peaks (this is the position at which the density profile becomes shallower
than $\rho(r)\propto r^{-2}$).  
Most of the mass has been stripped away by tidal forces and
now lies in two symmetric tidal tails of material that completely wrap
around the entire orbit.  In Fig.~\ref{f:008} we plot the evolution of the
circular velocity curve and density profile of this satellite.

\begin{figure}
\centering
\epsfysize=4.5truein
\epsffile{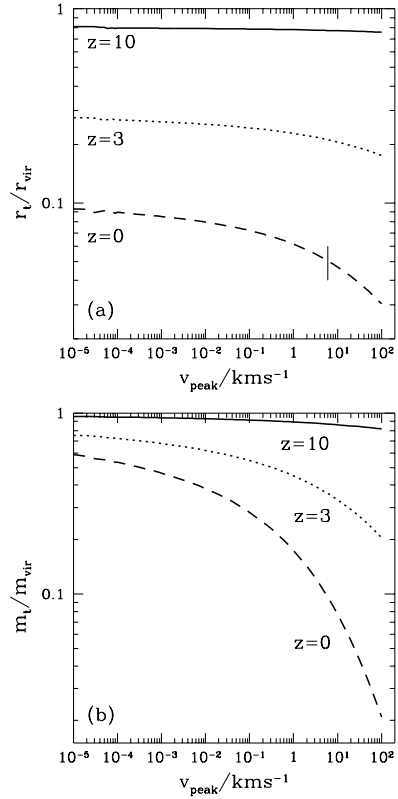}
\caption{The upper panel shows the ratio of tidal radius to virial 
radius for CDM subhalos of a given peak circular velocity orbiting at 
a distance of $R_\odot=8$ kpc from the center of a $10^{12}M_\odot$ CDM 
halo. We consider satellites with concentrations scaling from the fiducial 
$\Lambda$CDM model with structural parameters defined at three difference 
redshifts. We plot results down to halos with characteristic velocity of 1 
cm s$^{-1}$ which is roughly the free streaming limit for neutralinos. The 
lower panel shows the ratio of remaining bound mass to the initial virial 
masses of the subhalos. If subhalos enter Galactic progenitor halos at z=10 
then more than 90\% of their mass remains bound.}
\label{f:010}
\end{figure}

\noindent The
mass loss is continuous as particles on radial orbits escape, which in
turn decreases the tidal radius allowing more particles to escape.
Fig~\ref{f:009} shows the same satellite orbiting at 40 kpc within the
same potential.  In this case the tidal radius imposed by the Galaxy
is twice as large and lies beyond the inner core causing the
satellite to lose mass less rapidly. Whereas previously we found that
after 3 Gyrs only 0.3\% of the initial mass remains bound, in this
case 40\% remains bound at the final time.

Applying these results to subhalos orbiting within the Milky Way
allows us to estimate a radius at which a halo of a given
concentration will loose most of its mass to the smooth background,
i.e. the radius at which the Galactic tidal field truncates the
satellite at $r(v_{peak})$.  The ratio of $r_{peak}/r_{vir}$ decreases
with halo mass (halo concentration increases for small masses)
therefore smaller mass halos can survive intact deeper within the
Galactic potential. At the location of the earth within the Galactic
halo we estimate that halos with circular velocities larger than
$\approx 1$ km s$^{-1}$ {\it accreting today} will lose more than 95\% of
their mass.  Fig.~\ref{f:010} shows the tidal radii and masses of CDM subhalos
orbiting on circular orbits at
$R_\odot$ with a CDM parent halo of mass $10^{12}M_\odot$. 
To calculate tidal radii we use the orbital resonance theory derived by
Weinberg (1998) which is slightly more stringent than the standard
technique of using equipotential surfaces.
To make this calculation we have assumed that the
concentration $c=r_{vir}/2r_{peak}$  scales with mass as predicted by
NFW. Fig.~\ref{f:010} shows that the amount of tidally stripped mass depends 
sensitively on the redshift that subhalos accrete into the Milky Way.

\begin{figure}
\centering
\epsfxsize=\hsize\epsffile{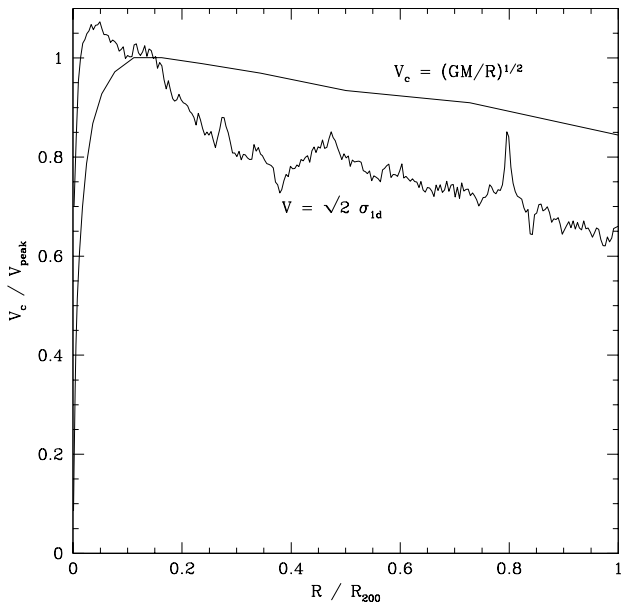}
\caption{The circular velocity profile,
($v_c=\sqrt{GM/R}$), and one dimensional velocity dispersion profile
(multiplied by $\sqrt{2}$) of one of the Local Group halos from
Section 2.}
\label{f:011}
\end{figure}

\section{Structure in velocity space}

The fraction of mass in the smooth component could vary from 0\% to
100\% depending solely on the subhalo structure, let alone the
formation history of the central Galactic halo.  If the halo is
predominantly smooth at $R_\odot$, as our current resolution limited
simulations find, then what can we learn by examining the velocity
distribution of dark matter particles?

\subsection{The global velocity dispersion profile}

We show the spherically averaged circular velocity, $v_c=\sqrt{GM/r}$,
in Fig.~\ref{f:011} along with the actual 1d velocity dispersion of particles
averaged in shells.  The velocity dispersion profile of Galactic mass
halos peaks sharply at about 3--7\% of the virial radius (approximately
$R_\odot$) and falls to a value of $\approx \sigma_{peak}/2$ at the halo
center and virial radius. 

\subsection{Anisotropy parameter}

Direct detection experiments are sensitive to the energy spectrum 
of dark matter particles arriving on the earth. Some detectors, such as DRIFT,
are also 
directionally sensitive to incoming WIMPS. The anisotropy
parameter can be defined as $\beta(r) = 1 - <v_t>^2/<v_r>^2$, where
$<v_r>$ and $<v_t>$ are the mean radial and tangential velocities at
radius $r$.  Using this definition, isotropic, radial and circular
orbits would have $\beta=0, 1, <0$ respectively.

\begin{figure}
\centering
\epsfxsize=\hsize\epsffile{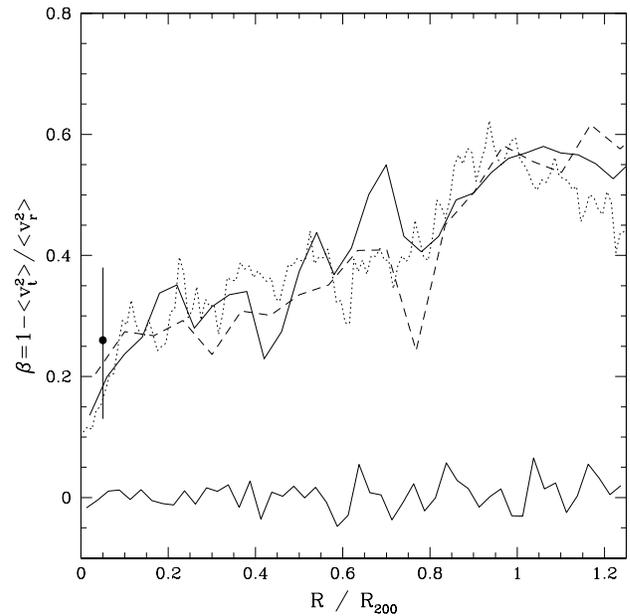}
\caption{The anisotropy parameter $\beta$ is plotted to the virial radii 
for a halo simulated at three different resolutions: dashed, solid and 
dotted are low, intermediate and high resolution simulations of the same 
halo.  The lower solid curve with $\beta \approx 0$ is a smooth halo 
constructed with similar structural parameters as the intermediate 
resolution CDM halo but with all particles on isotropic orbits, illustrating 
the Poisson noise in this quantity.  At the solar radius, $R_\odot \approx 
0.5\%R_{vir}$ we can expect values of $\beta=0.1-0.4$ from our limited set 
of six high resolution Galactic mass halos~ \protect\cite{moore99a} denoted 
by the single filled circle with an error bar.}
\label{f:012}
\end{figure}

In Fig.~\ref{f:012} we plot the anisotropy parameter of a single CDM halo
simulated with three different resolutions; $10^{4.3}, 10^{5.8}, 10^{6.7}$
particles within the virial radius and force resolution $<0.01R_{vir}$.
The value of
$\beta$ appears to be independent of resolution since we find the same
(noisy) trend in each simulation.  (We also found no change in shape
as we increase the numerical resolution -- the ratio of long:intermediate:short
axis in these three runs are the same to within 20\% at all radii.)
The behaviour of this halo is
typical of our simulations with $\beta$ rising from about 0.2 at
$R_\odot$ to 0.8 at $R_{vir}$. The average value of  $\beta\approx
0.26\pm0.1$  at $R_\odot=5\% R_{vir}$ was found using the results of 6
high resolution CDM halos from Ref. ~\cite{moore99b}.

\begin{figure}
\centering
\epsfxsize=\hsize\epsffile{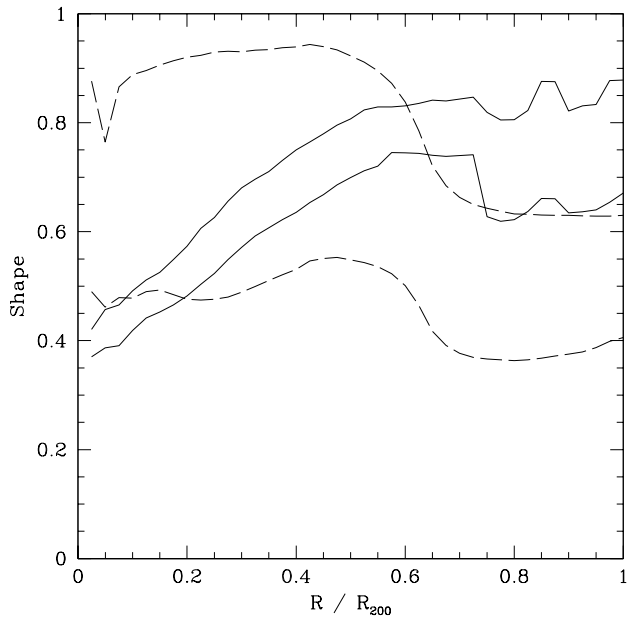}
\caption{The shapes of the two high resolution Local Group halos
plotted out to their virial radii. The upper and lower curves for each halo
show the intermediate:long and short:long axial rations.
One of the halos is prolate (solid
curves) and is highly flattened in the center but becomes more
spherical in the outer region.  The other halo is oblate (dashed
curves) but undergoes an abrupt shape change at about $0.5R_{vir}$ due
to a major accretion event.}
\label{f:013}
\end{figure}

\subsection{Anisotropy within triaxial halos}

\begin{figure}
\centering
\epsfxsize=\hsize\epsffile{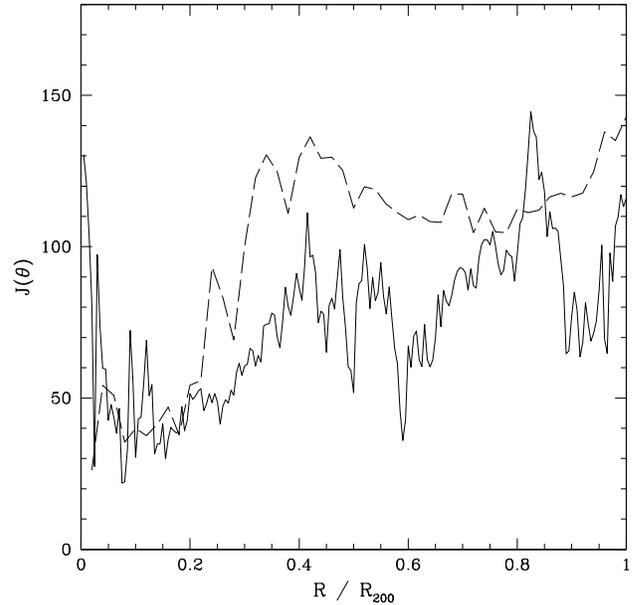}
\caption{The variation of the angular momentum vector as a function of
radius of the two halos shown in Fig.~\ref{f:013}. The direction of the
vector for the prolate halo (solid curve) wanders by about 50 degrees
whereas the oblate halo (dashed curve) changes sharply by 90 degrees
close to the region where the shape of the density distribution
changes.}
\label{f:014}
\end{figure}

Figure~\ref{f:013} shows the axial ratios of the two Local Group halos as a
function of radius. One of the halos (solid curves) is clearly
prolate, with short to long axis ratio similar to the intermediate to
long axis ratio. This halo is highly triaxial at the center but
becomes more spherical towards the virial radius.  The other halo
(dashed curve) is oblate in the center becoming more prolate in the
outer region. The origin of these shapes depend on the merger history
of the halos -- the oblate halo is particularly interesting since it
suffers a major merger at $z=0.7$. This leads to an abrupt change in
shape and angular momentum at $0.5r_{vir}$.
Figure~\ref{f:014} shows how the angle of the angular momentum vector
varies with radius from each of these  halo centers. The angular
momentum of shells varies with radius, wandering by $\gsim 100$
degrees from the center to the virial radius.  

CDM halos are clearly
complex systems with complicated formation histories.
They cannot be modelled as spherical or even simple triaxial
structures with constant flattening or angular momentum but they 
do have a remarkable regularity in their phase space density 
profiles ~\cite{taylor}. Although
axially symmetric collapse of a cold rotating mass distribution may 
give rise to significant caustic features ~\cite{sikivie99,green01}, 
this would not be a prediction of the CDM model.

\begin{figure}
\centering
\epsfxsize=\hsize\epsffile{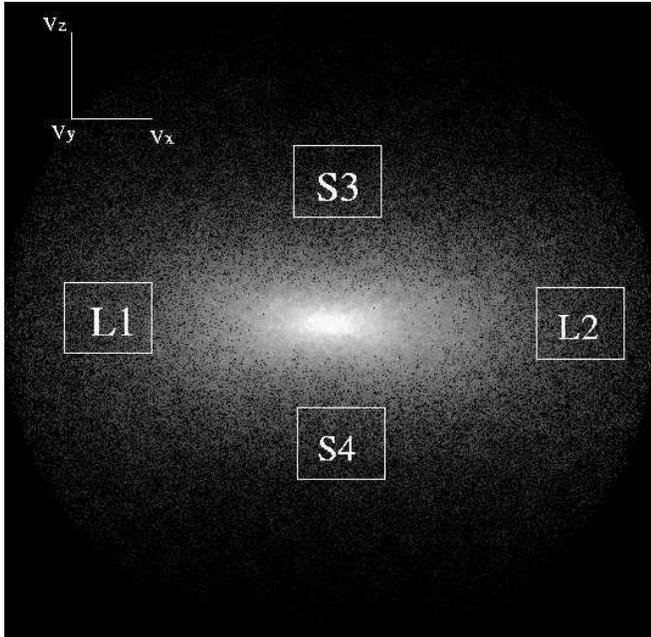}
\caption{The grey scale shows an edge on view of the particle distribution
within the inner 10\% of the prolate dark matter halo.
The boxes indicate cubes that contain 2000 particles that we use to
analyse the velocity anisotropy. L1 and L2 lie in the long axis of the
density distribution  whereas S3 and S4 lie in the short axis.}
\label{f:015}
\end{figure}

If a halo is triaxial then the energy distribution of particles will
provide the support for that shape (e.g., Ref.~\cite{evans00}). Here we
study the central velocity structure of the prolate halo in more
detail by sampling the particles in the small cubes indicated in
Fig.~\ref{f:015}.  We find a mean value of $\beta=0.36$ in the central 10\%
of the halo, whereas in positions L1 and L2, on the long axis, we
find that the orbits are even more radially biased with
$\beta=0.45$. On the short axis the orbits are still radially biased
however they are more circular than the mean with locations 
S3 and S4 having $\beta=0.25$.

\begin{figure}
\centering
\epsfxsize=\hsize\epsffile{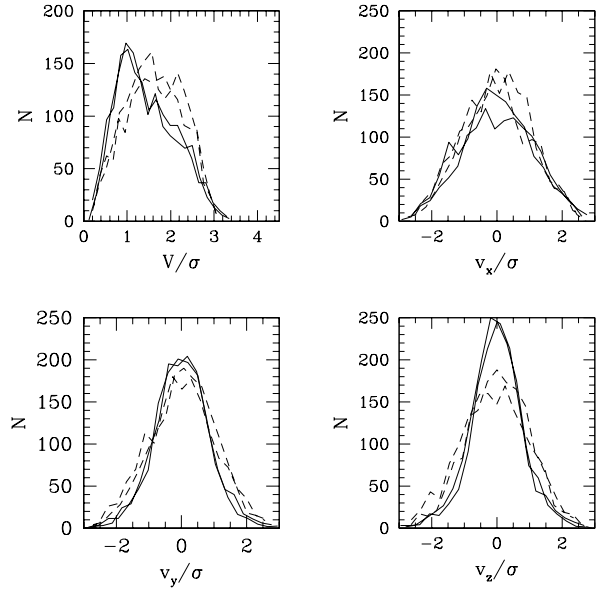}
\caption{The velocity histograms of particles taken from the regions
indicated in Fig.~\ref{f:015}.  The solid and dashed curves are in the long
and short axis respectively.  The first panel uses the total velocity
vector whereas the other three are just the $v_x, v_y$ and $v_z$
components. (The orientation of these vectors with respect to the
density distribution are shown in Fig.~\ref{f:015}.)}
\label{f:016}
\end{figure}

The distribution of particle velocities and speeds along the short and
long axis are shown in Fig.~\ref{f:016}. The height and width of the velocity
distribution varies significantly and a dark matter 
detector may find different event rates and signal profiles depending
on its location.
The differential rate for detecting CDM particles is simply
$$
{dR\over dQ} \propto \int^{\inf}_{v_{min}} {f_r(v)\over v} dv
$$
where $R$ is the rate, $Q$ is the recoil energy and $f_r(v)$ is the
probability distribution of particle speeds relative to the detector
~\cite{jungman96}.
In a forthcoming paper we will make predictions for event rates and
modulation signals using the velocity distributions obtained directly
from these halo simulations.

\subsection{Dark matter streams}

A halo that accretes a
smaller mass satellite will tidally disrupt the satellite into tidal
tails that slowly wrap in phase space (c.f., Ref.~\cite{helmi99,stiff01}). 
The symmetric tidal tails in Fig.~\ref{f:007} illustrated this process for 
a single high resolution satellite. The presence of dark matter streams
may be inferred using directional dark matter detectors such as DRIFT,
or by finding a highly peaked signal resulting from particles entering the
detector with similar energies.

Numerical resolution sets a limit to the masses of the first clumps of
dark matter that collapse within the CDM model. Most of the mass that
has collapsed by z=10 ends up making the inner few percent of the
final halo.  The first clumps to collapse are extremely dense yet have
only a handful of particles - thus they have a smooth phase space
structure that may be dominated by relaxation effects. In order to
correctly follow the phase space structure at $R_\odot$ it is
necessary to correctly resolve the formation of the progenitor halos
at $z=10$. This would require a simulation with at least $10^{10}$
particles.

\begin{figure}
\centering
\epsfxsize=\hsize
\epsffile{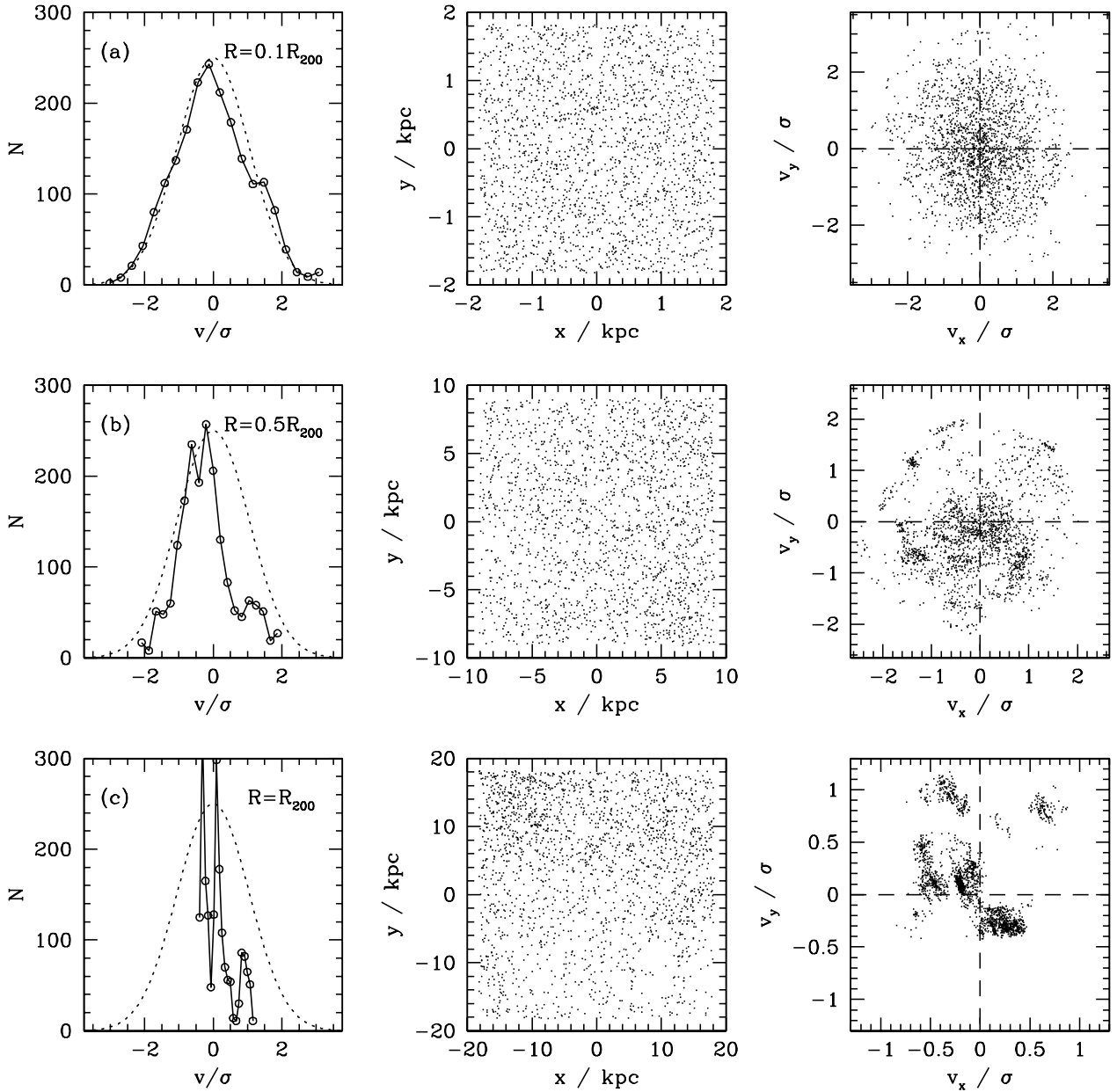}
\caption{The structure in velocity space as a function of position
within the simulated Galactic halo. We have taken 3 cubes at different
positions in the halo that are smooth in density space. Each cube
contains about 1000 particles and has no bound subhalos within them as can
been seen in the central panels.  The left panel shows the histogram
of one component of the distribution of particle speeds in each cube
-- the curve shows a Maxwellian distribution with velocity dispersion
$\sigma$ set equal to the spherically averaged value at
$0.1R_{vir}$. The right panels show the $x$ and $y$ components of
velocity plotted against each other.}
\label{f:017}
\end{figure}

Figure~\ref{f:017} shows the velocity structure within one of the Local Group
CDM halos sampled at three different radii -- close to the solar
radius, at half the virial radius and near the edge of the virial
radius. The amount of structure in velocity space depends on where we
sample the halo. In the outer region we find that all of the particles
lie within several distinct streams
with a velocity distribution that is remarkably non-Gaussian. At the
halo center, the streams have phase wrapped several times and give
rise to a smooth structure in velocity space.

The key question that we would like to answer is;  {\it on a scale of
order the size of the earths orbit around the sun, are we dominated by
a single stream of dark matter, or by many thousands of overlapping
streams?}

Our simulation shows that the phase space distribution is smooth at $R_\odot$.
However we can only resolve subhalos as small as $10^8M_\odot$ which have
internal velocity dispersions of order 10 km s$^{-1}$ and can phase 
wrap several
times around the Galaxy in a Hubble time. Therefore in our
patch of phase space that is 2kpc on a side, we find multiple streams
from the same tidally disrupted halos. 

\twocolumn[\hsize\textwidth\columnwidth\hsize\csname 
@twocolumnfalse\endcsname
 ]

\noindent If we had infinite resolution and examined a volume that was 
one parsec on a side what would we find? We would then be sensitive to the 
substructure within our poorly resolved $10^8M_\odot$ halos.

A tiny CDM halo with $v_{peak}=1$ m s$^{-1}$ has $M_{vir}=10^{-6}M_\odot$,
$r_{vir}=0.5$ parsecs, $r_{peak}=0.001$ parsecs and a concentration c=400 at a
redshift z=0.  Its tidal radius at $R_\odot$ is $r_t=0.05$ parsecs which
encloses a mass of $10^{-7}M_\odot$. A straightforward simulation of this
structure orbiting at $R_\odot$ demonstrates that its tidal streams would be
$\approx 50$ parsecs long and 0.1 parsecs wide -- certainly not phase wrapped.
A halo with the same circular velocity at a redshift z=10 would have even smaller
tidal streams. The mean mass density
of CDM at $R_\odot$ is $\sim 10^{-3}M_\odot/{\rm pc}^3$, therefore if the
central halo was made entirely 
from streams of tidally disrupted 1 m s$^{-1}$ halos at z=10, 
we estimate that their filling factor is of order unity, with large 
uncertainties. 

Future analytic work is clearly necessary to resolve these issues. However, 
it is possible that the density of dark matter in the solar system is zero,
or that the solar system may be moving through a single stream of CDM 
particles that has bulk motion of $\approx 200$ km s$^{-1}$, with a 
velocity distribution of only a few m s$^{-1}$.

\section{Conclusions}

We have carried out a high resolution numerical
simulation of the Local Group and one of its substructure halos
focussing our analysis on properties relevant 
to direct detection experiments. We summarise our main conclusions here:

$\bullet$ 
We combined our halo density profiles with observational
constraints on the baryonic content of the Galaxy to infer the local
density of dark matter.
Observations favour a dominant baryonic component within $R_\odot=8$ kpc
which leaves a CDM halo that contributes
$\lsim 3 \times 10^{10}M_\odot$ within the same radius. 
This constrains the local dark matter density to be $\rho(R_\odot)\lsim 0.23$
GeV/cm$^3$ for a Galaxy with $v_\odot=220$ km s$^{-1}$.

$\bullet$  
CDM halos form via a series of many mergers and
accretion events. The final triaxial halos have a complex structure; the
shapes can vary from oblate to prolate and the angular momentum axis
can change by more than 100 degrees in a single system.  Axially
symmetric collapse can not be applied to infer the structure of CDM
halos -- neither do we find or expect to find significant caustic
features, such as rings or shells.

$\bullet$  
We increased the resolution within a region that collapses to form a
$10^8M_\odot$ halo that is typical of those that may surround the
Galactic dwarf spheroidal galaxies. Even on this mass scale
CDM halos have cuspy singular density
profiles and contain a remarkable
amount of surviving substructure -- dwarf spheroidals resemble
scaled versions of galaxies or clusters. The slope
of the circular velocity function is steeper in higher mass halos,
perhaps indicating a dependence on the slope of the power spectrum.

$\bullet$ 
The smallest dark matter halos that we can resolve, with circular velocities
$\sim 1$ km s$^{-1}$, have cuspy singular density profiles and concentration 
parameters that scale as predicted by Navarro {\it et al.}~\cite{navarro96}.

$\bullet$
We show that halos with circular velocities $\lsim 1$ km s$^{-1}$ can survive
orbiting at the solar position within the Galactic potential. Since the
central Galactic halo is in place by z=10, nearly all of the accreted halos
will remain intact and retain most of their mass.
The presence of a smooth component of dark matter at $R_\odot$ depends on
the detailed merger history of the Galaxy and on the internal
structure of the first CDM halos to collapse with characteristic
velocities of 1 m s$^{-1}$ -- 1 km s$^{-1}$.

$\bullet$  
The orbits of the particles in the smooth component are
radially biased -- the spherically averaged anisotropy parameter at
$R_\odot$ may lie in the range $0.1\lsim \beta \lsim 0.4$. 

$\bullet$ 
The velocity distribution of CDM particles varies significantly 
with the location of the observer and cannot be approximated by a
single Maxwellian.

$\bullet$  
At our current resolution, the outer halo is made entirely
from discrete streams of material originating from tidally stripped
subhalos. Closer to the halo centre these streams have phased wrapped
several times. Further analytic and numerical work is required to
investigate the numbers of streams flowing through the solar system or
the fraction of dark matter that is smoothly distributed on this scale.

\section*{Acknowledgements}

Many thanks to Carlos Frenk, Amina Helmi, Adrian Jenkins,
Luis Theodoro and Simon White for useful discussions. Ben Moore 
is supported by the Royal Society and acknowledges the Virgo 
Consortium for computing resources.  Carlos Calc\'{a}neo-Rold\'{a}n 
continues his research thanks to a grant from CONACyT (Mexico).


\begin{references}

\bibitem{moore94} B. Moore, Nature, {\bf 370}, 620 (1994).

\bibitem{flores94} R.A. Flores, and J. R. Primack, Astrophys. J. Lett. 
{\bf 457}, L5 (1994).

\bibitem{burkert95} A. Burkert, Astrophys. J. Lett., {\bf 447}, L25  (1995). 

\bibitem{klypin99a} A. Klypin, A. V. Kravtsov, O. Valenzuela, and F. Prada, 
Astrophys. J. {\bf 522}, 82 (1999).

\bibitem{moore99a} B. Moore, S. Ghigna, F. Governato, G. Lake, T. Quinn, 
J. Stadel, and P. Tozzi, Astrophys. J. Lett. {\bf 524}, L19 (1999).

\bibitem{sellwood00} J. A. Sellwood, Astrophys. J. {\bf 540}, L1 (2000).

\bibitem{berg98} L. Bergstr{\"o}m, New Astron. Rev., {\bf 42}, 245 (1998).

\bibitem{griest88} K. Griest, Phys. Rev. Lett. {\bf 61}, 666 (1988).

\bibitem{DRIFT99}
M.~J.~Lehner, K.~Griest, C.~J.~Martoff, G.~E.~Masek, T.~Ohnuki, 
D.~Snowden-Ifft and N.~J.~Spooner,
astro-ph/9905074.

\bibitem{DAMA00} R.~Bernabei {\it et al.}  [DAMA Collaboration],
Phys. Lett. B {\bf 480}, 23 (2000).

\bibitem{CDMS00} R.~Abusaidi {\it et al.}  [CDMS Collaboration],
Phys. Rev. Lett. {\bf 84}, 5699 (2000).

\bibitem{EDELWEISS} A.~Benoit {\it et al.}  [EDELWEISS Collaboration],
astro-ph/0106094.

\bibitem{copi01} C. J. Copi, and L. M. Krauss, Phys. Rev. D. {\bf 63} 
043507 (2001). 

\bibitem{ullio00} P.~Ullio and M.~Kamionkowski,
JHEP {\bf 0103}, 049 (2001).

\bibitem{widrow00} L. M. Widrow, Astrophys. J., Suppl. Ser. 
{\bf 131}, 39 (2000).

\bibitem{stiff01} D. Stiff, L. M. Widrow, J. Frieman,
astro-ph/0106048.

\bibitem{calca00} C. Calc\'aneo-Rold\'an, and B. Moore, Phys. Rev. D. 
{\bf 62} 123005 (2000).

\bibitem{brhlik99} M.~Brhlik and L.~Roszkowski,
Phys. Lett. B {\bf 464}, 303 (1999)

\bibitem{kamio98} M. Kamionkowski, and A. Kinkhabwala, Phys. Rev. D. 
{\bf 57}, 3256 (1998).

\bibitem{evans00} N. W. Evans, C. M. Carollo, C.M. and P. T. de Zeeuw, 
Mon. Not. R. Astron. Soc. {\bf 318}, 1131 (2000). 

\bibitem{sikivie99} P. Sikivie, Phys. Rev. D {\bf 60}, 063501 (1999).

\bibitem{moore01} B.~Moore,
astro-ph/0103094.

\bibitem{ghigna98} S. Ghigna, B. Moore, F. Governato, G. Lake, T. Quinn, and 
J. Stadel, Mon. Not. R. Astron. Soc. {\bf 300}, 146 (1998).

\bibitem{schwarz00} D. J. Schwarz, and S. Hofmann, Nucl. Phys. Proc. Suppl. 
{\bf 87}, 93 (2000).

\bibitem{schmid97} C. Schmid, D. J. Schwarz, and P. Widerin, 
Phys. Rev. Lett. {\bf 78}, 791 (1997).

\bibitem{white76} S. D. M. White, Mon. Not. R. Astron. Soc. 
{\bf 177}, 717 (1976).

\bibitem{white78} S. D. M. White and M. Rees, Mon. Not. R. Astron. Soc. 
{\bf 183}, 341 (1978).

\bibitem{moore96} B. Moore, N. Katz, and G. Lake, Astrophys. J. {\bf 457}, 
455 (1996).

\bibitem{moore98} B. Moore, F. Governato, T. Quinn, J. Stadel, and G. Lake, 
Astrophys. J. {\bf 499}, L5 (1998).

\bibitem{moore99b} B. Moore, T. Quinn, F. Governato, J. Stadel, and G. Lake 
Mon. Not. R. Astron. Soc. {\bf 310}, 1147 (1999).

\bibitem{klypin99b} A. Klypin, S. Gottl\"ober, A. Kravtsov, and 
M. Khokhlov, Astrophys. J. {\bf 516}, 530 (1999b).

\bibitem{governato97} F. Governato B. Moore, R. Cen, J. Stadel, G. Lake and 
T. Quinn New Astron. {\bf 2}, 91 (1997).

\bibitem{navarro96} J. F. Navarro, C. S. Frenk, and S. D. M. White, 
Astrophys. J. {\bf 462}, 563 (1996).

\bibitem{jang01} H. Jang-Condell, and L. Hernquist, Astrophys. J. 
{\bf 548}, 68 (2001)

\bibitem{freese88} K. Freese, J. Frieman, and A. Gould, Phys. Rev. D. 
{\bf 37}, 3388 (1988).          

\bibitem{eke01} V. R. Eke, J. F. Navarro, and M. Steinmetz, 
astro-ph/0012337. 

\bibitem{sackett97} P. D. Sackett, Astrophys. J. {\bf 483}, 103 (1997).

\bibitem{kuijken91} K. Kuijken K. and G. Gilmore G. Astrophys. J. Lett. 
{\bf 367}, L9 (1991).

\bibitem{flynn94} C. Flynn, and B. Fuchs, Mon. Not. R. Astron. Soc. 
{\bf 270}, 471 (1994).

\bibitem{flynn99} C. Flynn, A. Gould, and J. Bahcall, 
Astron. Soc. Pac. Conf. Ser. {\bf 165}, 387 (1999).

\bibitem{dehnen98} W. Dehnen, and J. Binney, Mon. Not. R. Astron. Soc. 
{\bf 298}, 387 (1998).

\bibitem{gerhard99} O. E. Gerhard, Astron. Soc. Pac. Conf. Ser. 
{\bf 182}, 307 (1999). 

\bibitem{binney00} J. Binney, N. Bissantz, N., and O. Gerhard, 
Astrophys. J. Lett. {\bf 537} L99 (2000). 

\bibitem{gerhard00} O. Gerhard,  Astron. Soc. Pac. Conf. Ser. 
{\bf 197}, 201 (2000).

\bibitem{debatt98} V. P. Debattista and J. A. Sellwood Astrophys. J. Lett. 
{\bf 493}, L5 (1998). 

\bibitem{weiner01} B. J. Weiner, J. A. Sellwood, and T. B. Williams, 
Astrophys. J. {\bf 546}, 931 (2001). 

\bibitem{white93} S. D. M. White, J. F. Navarro, A. E. Evrard, and 
C. S. Frenk, Nature (London) {\bf 366} 429 (1993).

\bibitem{navarro00} J. F. Navarro, and M. Steinmetz, Astrophys. J. 
{\bf 528}, 607 (2000).

\bibitem{tytler99} D. Tytler, S. Burles, L. Lu, X. Fan, A. Wolfe, and 
B. D. Savage, Astron. J. {\bf 117}, 63 (1999). 

\bibitem{wilkinson99} M. I. Wilkinson, and N. W. Evans, 
Mon. Not. R. Astron. Soc. {\bf 310}, 645 (1999). 

\bibitem{ibata00} R. Ibata, G. F. Lewis, M. Irwin, E. Totten, and 
T. Quinn, Astrophys. J. {\bf 551}, 294 (2001).

\bibitem{olling2001} R.~Olling, and M.~R.~Merrifield, astro-ph/0104465.

\bibitem{barnes84} J.~Barnes and S.~D.~M. White, Mon. Not. R. Astron. Soc. 
{\bf 211}, 753 (1984).  

\bibitem{peebles70} P. J. E. Peebles, Astron. J. {\bf 75}, 13 (1970).

\bibitem{taylor} J. E. Taylor, and J. F. Navarro, 
astro-ph/0104002.

\bibitem{green01} A. M. Green, Phys. Rev. D. {\bf 63 }, 103003 (2001).

\bibitem{jungman96} G. Jungman, M. Kamionkowski, and K. Griest, Phys. Rep. 
{\bf 267}, 195 (1996).   

\bibitem{helmi99} A. Helmi, and S. D. M. White, Mon. Not. R. Astron. Soc. 
{\bf 307}, 495. (1999).

\end{references}
\end{document}